\documentclass[10pt,leqno]{amsart}

\usepackage[a4paper,margin=1.3in]{geometry}

\usepackage{xcolor}
\usepackage{graphicx}
\usepackage{algorithm}
\usepackage{bm}
\usepackage{indentfirst,csquotes}
\usepackage{amssymb,amsthm,amsmath}
\usepackage{paralist,fancyhdr,etoolbox}
\usepackage{caption}
\usepackage{lineno}

\makeatletter
\renewcommand{\subsection}{\@startsection
  {subsection}
  {2}
  {\z@}
  {-2.5ex \@plus -1ex \@minus -.2ex}
  {1.5ex \@plus .2ex}
  {\normalfont\centering\itshape}
}
\makeatother

\usepackage[colorlinks=true,linkcolor=blue,citecolor=blue,urlcolor=black]{hyperref}

\newtheorem{theorem}{Theorem}

\newtheorem{lemma}{Lemma}
\newtheorem{proposition}{Proposition}

\theoremstyle{definition}

\usepackage{lipsum}

\newcommand{\DOI}[1]{\href{https://doi.org/#1}{doi:#1}}

\begin{document}
\setlength{\abovedisplayskip}{10pt}
\setlength{\belowdisplayskip}{10pt}
\setlength{\abovedisplayshortskip}{6pt}
\setlength{\belowdisplayshortskip}{10pt}

\title{A test for normality based on self-similarity} 

\maketitle

\begin{center}

{\scshape Akin Anarat}\\[4pt]
{\itshape Mathematical Institute, Heinrich Heine University Düsseldorf}\\[2pt]
\texttt{akin.anarat@hhu.de}

\vspace{1.2em}

{\scshape Holger Schwender}\\[4pt]
{\itshape Mathematical Institute, Heinrich Heine University Düsseldorf}\\[2pt]
\texttt{holger.schwender@hhu.de}
\end{center}


\let\thefootnote\relax

\begin{abstract}
Testing for normality is a widely used procedure in statistics and data analysis, often applied prior to employing methods that rely on the assumption of normally distributed data. While several existing tests target distributional characteristics such as higher-order moments, others focus on functional aspects such as the distribution function of the normal distribution.

In this article, we propose an alternative idea to these common approaches by exploiting the self-similarity property of the normal distribution and introduce the Self-Similarity Test for Normality (SSTN). This testing procedure leverages the structural property that the distribution of a suitably centered and scaled sum of independent and identically distributed random variables with finite variance coincides with the original distribution if and only if that distribution is normal.

The SSTN evaluates normality by applying a self-similarity transformation to the standardized empirical characteristic function and examining how the transformed functions change across successive applications of this transformation. For the normal distribution, repeated applications of this transformation preserve the functional form of the characteristic function, whereas deviation from normality manifests in systematic changes between consecutive transforms. The magnitude of these changes is aggregated into a test statistic, whose null distribution is obtained by Monte Carlo calibration, using a sample-size-specific calibration for small samples and a Monte Carlo approximation of the asymptotic null distribution for larger ones.

We perform a comprehensive simulation study considering a wide range of simulation scenarios to evaluate the SSTN and compare it with several well-established tests for normality. In this simulation study, the SSTN shows a performance that is at least competitive and frequently superior to the performances of the other tests.
\end{abstract} 

\begin{center}
\begin{minipage}{.82\textwidth}
\setlength{\baselineskip}{0.8\baselineskip}%
\end{minipage}
\end{center}

\section{Introduction}

Many statistical methods across various fields of application rely on the assumption that the data under analysis follow a normal distribution. This assumption can be particularly important, as the normal distribution exhibits several convenient properties, such as symmetry, finite moments of all orders, and closure under affine transformations. These properties ensure the validity of various inferential procedures. Across statistical techniques such as parametric estimation, hypothesis testing, and model-based prediction, assessing the plausibility of normality is, therefore, often an essential initial step.

Numerous methods have been developed to investigate whether observed data originate from a normal distribution. Within the class of tests for normality, a wide variety of methodological approaches can be distinguished. Some approaches target specific distributional characteristics like asymmetry or heavy tails by analyzing higher-order moments, while others assess functional aspects of the underlying distribution \cite{thode2002}.

Among the moment-based approaches, the Jarque–Bera test \cite{jarque1980} assesses departures from normality by evaluating skewness and kurtosis jointly, while the D’Agostino–Pearson test \cite{dagostino1973} combines separate measures of skewness and kurtosis into a single omnibus statistic. Within the class of function-based procedures, e.g., the Shapiro–Wilk test \cite{shapiro1965} relies on correlations between ordered sample values and their expected normal counterparts, the Anderson–Darling test \cite{anderson1954} places particular emphasis on deviations between empirical and theoretical distribution functions, and the Lilliefors test \cite{lilliefors1967} adapts the Kolmogorov–Smirnov framework \cite{smirnov1948} to the case of testing normality when the mean and variance are estimated from the data.

However, none of these established procedures exploits a distinctive structural property of the normal distribution, namely its self-similarity under convolution \cite{samorodnitsky2016}, where self-similarity means that the distribution of a suitably centered and scaled sum of arbitrarily many independent and identically distributed random variables coincides with the distribution of each summand. It is well-known that under finite variance this property holds if and only if the underlying distribution is normal. Hence, among all distributions with finite variance, the normal distribution is uniquely characterized by the fact that it remains invariant under convolution and rescaling. This yields a conceptually different approach to testing for normality, as it targets a defining global structural property of the normal distribution rather than specific distributional features.

This property of self-similarity can equivalently be expressed on the level of characteristic functions, which uniquely determine probability distributions. In particular, the characteristic function of a suitably centered and scaled sum of i.i.d.\ random variables with finite variance coincides with that of the summands themselves if and only if the underlying distribution is normal \cite{kallenberg2021}.

In this manuscript, we introduce the Self-Similarity Test for Normality (SSTN), a new statistical test that essentially makes use of this structural property of self-similarity by assessing it through a standardized version of the empirical characteristic function of the observed data. Several tests for normality or goodness-of-fit are based on empirical characteristic functions \cite{meintanis2016}. In contrast to such tests, which typically rely on discrepancies between empirical and theoretical characteristic functions \cite{epps1983, murota1981} or between different empirical versions \cite{alba2001}, the SSTN directly exploits the self-similarity property of the normal distribution. In particular, the SSTN employs a standardized empirical characteristic function, similarly to the studentized empirical characteristic function considered by Murota and Takeuchi \cite{murota1981}, but uses it to assess the self-similarity structure that characterizes normality rather than comparing it directly to the characteristic function of the normal distribution.

While self-similarity is defined in terms of the distribution of scaled and shifted sums of i.i.d.\ random variables, it can be evaluated in practice from a single sample by applying an appropriate transformation directly to its empirical characteristic function. Following this idea, the SSTN makes use of this self-similarity property of the normal distribution to test a set of observations for normality by iteratively applying a self-similarity transformation to the standardized empirical characteristic function and aggregating the discrepancies between consecutive iterates of these estimated characteristic functions up to a fixed iteration number. For inference, the null distribution of the test statistic is obtained through a calibration procedure under the null hypothesis of normality.

The remainder of this article is structured as follows. In Section~\ref{s:SSTN}, the SSTN is introduced and its theoretical framework is developed. Section~\ref{s:Numerical_Implementation} describes the numerical procedure of the SSTN based on this framework. In Section~\ref{s:Simulation_study}, a comprehensive simulation study is conducted to compare the SSTN to several well-established tests for normality. Finally, Section~\ref{s:Conclusions} concludes this article.

\section{Self-Similarity Test for Normality} \label{s:SSTN}

\subsection{Characterization of normality via self-similarity}\label{s:self-similarity}

Self-similarity can formally be defined as follows. A distribution $P_X$ on $\mathbb{R}$ is called self-similar if, for every $m \in \mathbb{N}$, there exist constants $a_m > 0$ and $b_m \in \mathbb{R}$ such that
\begin{equation}\label{f:self-similarity}
a_m \sum_{j=1}^m X_j + b_m \overset{\text{d}}{=} X,
\end{equation}
where $X_1, \dots, X_m$ are i.i.d.\ copies of a random variable $X$ with distribution $P_X$. The normal distribution $\mathcal{N}(\mu,\sigma^2)$ satisfies \eqref{f:self-similarity} with $a_m = 1/\sqrt{m}$ and $b_m = (1-\sqrt{m})\mu$ and is the only self-similar distribution among all non-degenerate distributions with finite variance \cite{kallenberg2021, nolan2020}. This uniqueness forms the conceptual basis of the SSTN.

Because characteristic functions uniquely determine probability distributions, the self-similarity property can equivalently be expressed in terms of characteristic functions. For a real-valued random variable $X$, the characteristic function is defined as 
\[
\psi_X(t) = \mathbb E[\exp(itX)], \qquad t \in \mathbb{R}.
\]
Using the properties of characteristic functions for linear transformations and sums of independent random variables, \eqref{f:self-similarity} can equivalently be expressed as
\begin{equation}\label{f:self-similarity_cf}
\exp\!\left(itb_m\right)\bigl(\psi_X(a_m t)\bigr)^m = \psi_X(t),
\qquad t \in \mathbb{R}.
\end{equation}
For a normally distributed random variable $X \sim \mathcal{N}(\mu,\sigma^2)$, inserting $a_m = 1/\sqrt{m}$ and $b_m = (1-\sqrt{m})\mu$ into \eqref{f:self-similarity_cf} yields the identity
\begin{equation}\label{f:self-similarity_cf_normal}
\exp\!\left(it(1-\sqrt{m})\mu\right) \left(\psi_X\!\left(\frac{t}{\sqrt{m}}\right)\right)^m
= \psi_X(t) \qquad \text{for all } m \in \mathbb{N}.
\end{equation}

Since the formulation in \eqref{f:self-similarity} involves sums of independent copies of a random variable $X$, verifying this relation directly would, in practice, require access to independent realizations of such sums. In contrast, the identity in \eqref{f:self-similarity_cf_normal} reformulates self-similarity on the level of characteristic functions, whose empirical counterparts can be estimated from a single sample of i.i.d.\ observations. This functional perspective forms the foundation of the SSTN, as the characteristic function of a sum of independent random variables equals the product of their characteristic functions, implying that all finite convolution powers of the underlying distribution are implicitly encoded in the characteristic function.

\subsection{Test statistic of the SSTN} \label{s:Test_Statistic_SSTN}

In this section, we introduce the construction of the test statistic of the SSTN by detailing how the empirical characteristic function is used to assess self-similarity and by defining the resulting test statistic.

In the following, $n$ denotes the sample size of the observed realizations of i.i.d. random variables $X_1,\dots,X_n$ with distribution $P_X$, whereas $m$ is a conceptual parameter that refers to the number of hypothetical summands of independent copies of the standardized distribution of $X_1,\dots,X_n$ used to verify self-similarity.

As usually considered in tests for normality, the SSTN tests the hypothesis
\[
\begin{array}{c}
H_0\colon\ P_X = \mathcal{N}(\mu,\sigma^2)\ 
\quad \text{for some } \mu\in\mathbb{R},\ \sigma^2>0, \\[1.5ex]
\text{vs.} \\[1.5ex]
H_1\colon\ P_X \neq \mathcal{N}(\mu,\sigma^2)\ 
\quad \text{for all } \mu\in\mathbb{R},\ \sigma^2>0.
\end{array}
\]

To evaluate this hypothesis, the SSTN utilizes the empirical characteristic function
\[
\widehat{\psi}_X(t)
=
\frac{1}{n}\sum_{k=1}^n \exp(itX_k), 
\qquad t\in\mathbb{R},
\]
as an estimator of the true (unknown) characteristic function $\psi_X$. To account for the unknown mean and variance of $P_X$, the SSTN employs the standardized version
\begin{equation}\label{f:standardized_ecf}
\widehat{\psi}_X^*(t)
=
\exp\!\left(-it\,\frac{\bar X}{S_X}\right)
\widehat{\psi}_X\!\left(\frac{t}{S_X}\right),
\qquad t\in\mathbb{R},
\end{equation}
of $\widehat{\psi}_X(t)$, where
\[
\bar X = \frac{1}{n}\sum_{k=1}^n X_k
\qquad \text{and} \qquad
S_X^2 = \frac{1}{n}\sum_{k=1}^n (X_k - \bar X)^2
\]
denote the sample mean and the sample variance, respectively, where the normalization by $1/n$ in the definition of the sample variance is employed to simplify subsequent algebraic derivations. Under the null hypothesis, $\widehat{\psi}_X^*(t)$ estimates the characteristic function $\psi_0(t) = \exp(-t^2/2)$ of the standard normal distribution.

To connect the structural identity in \eqref{f:self-similarity_cf_normal} with an empirical procedure, we first formalize the corresponding functional transformation. For $m\in\mathbb{N}$ and any $f \in C(\mathbb{R};\mathbb{C})$, i.e.\ any continuous function from $\mathbb{R}$ to $\mathbb{C}$, define the self-similarity operator by
\[
\mathcal S_m[f](t)
:=
\left(f\!\left(\frac{t}{\sqrt{m}}\right)\right)^{m},
\qquad t \in \mathbb{R}.
\]
In the SSTN, this operator is used to specify the empirical self-similarity transform $S_m^{\!(n)}$ for $m\in\mathbb{N}$ by
\begin{equation}
\label{f:self-similarity_transform}
S_m^{\!(n)}(t)
:=
\mathcal S_m\!\left[\widehat{\psi}_X^*\right](t)
=
\left(\widehat{\psi}_X^*\left(\frac{t}{\sqrt{m}}\right)\right)^{m},
\qquad t \in \mathbb{R}.
\end{equation}

To assess deviations from self-similarity, the SSTN examines how strongly the functions $S_1^{\!(n)}, \dots, S_{M+1}^{\!(n)}$ differ from each other. To this end, the iterative discrepancies
\begin{equation}
\label{f:iterative_discrepancies}
D_m^{\!(n)}(t) := S_{m+1}^{\!(n)}(t) - S_m^{\!(n)}(t),
\qquad m = 1, \dots, M,
\end{equation}
are determined for a fixed number $M \in \mathbb{N}$.

Considering these iterative discrepancies is motivated by the fact that among all distributions with finite variance only the characteristic function $\psi_0$ of the standard normal distribution satisfies $\mathcal S_m[\psi_0](t) = \psi_0(t)$ for all $m \in \mathbb{N}$, and therefore, 
\[
\mathcal S_{m+1}[\psi_0](t) = \mathcal S_m[\psi_0](t) \quad \text{for all } m \in \mathbb{N},\ t \in \mathbb{R}.
\] 
Since $\widehat{\psi}_X^*$ serves as an estimator of $\psi_0$ under the null hypothesis that $P_X$ is a normal distribution, the iterative discrepancies $D_m^{\!(n)}$ are expected to be small for all $m \in \mathbb{N}$, when the considered data stems from a normal distribution, whereas deviations from normality induce larger discrepancies.

To obtain a scalar measure of these discrepancies, the self-similarity discrepancy measure
\begin{equation}
\label{f:self-similarity_discrepancy_measures}
\Delta_m^{\!(n)}
:=
\int_{-T}^T w(t)\, \big|D_m^{\!(n)}(t)\big|^2\, dt,
\qquad m = 1, \dots, M,
\end{equation}
is considered for a fixed number $T > 0$, where the choice of $T$ is discussed in Section~\ref{s:Numerical_Implementation} and $w$ is a real-valued and bounded weighting function that determines the contribution of different values of $t \in [-T, T]$. In the SSTN, we employ the weighting function
\begin{equation}\label{f:weighting}
w(t) = \exp\left(- \beta t^2\right)
\end{equation}
where $\beta > 0$ is a tuning parameter. This weighting function is motivated by the observation of Epps \cite{epps2005} that weighting functions of this form are commonly used in tests for normality based on empirical characteristic functions.

After determining the measures $\Delta_m^{\!(n)}$, these quantities are standardized to obtain the standardized discrepancies
\begin{equation}
\label{f:standardized_discrepancies}
T_m^{\!(n)} 
:= \frac{\Delta_m^{\!(n)} - \mathbb E_{H_0}\bigl[\Delta_m^{\!(n)}\bigr]}{\sqrt{\mathrm{Var}_{H_0}\bigl(\Delta_m^{\!(n)}\bigr)}},
\qquad m=1,\dots,M,
\end{equation}
where the mean and the variance under the null hypothesis are taken with respect to the distribution of $\Delta_m^{\!(n)}$ induced by the standard normal model.

This standardization is essential in the SSTN because all iteration levels $m=1,\dots,M$ are taken into account in the construction of the SSTN test statistic, while the raw discrepancy measures $\Delta_m^{\!(n)}$ are not directly comparable across different values of $m$. As shown in Lemma~S1 of the Supplementary Material, the empirical self-similarity transform satisfies
\[
S_m^{\!(n)}(t)\to \psi_0(t)
\qquad\text{for all } t\in\mathbb R
\]
as $m\to\infty$ for every non-degenerate underlying distribution, including those without finite variance. Consequently, the discrepancy measures $\Delta_m^{\!(n)}$ are expected to become small for sufficiently large values of $m$ not only under the null hypothesis, but also for non-normal distributions.

By centering and scaling $\Delta_m^{\!(n)}$ with respect to its expectation and variance under the null hypothesis, the resulting standardized statistics $T_m^{\!(n)}$ assess each discrepancy relative to its null distribution at the corresponding iteration level. This prevents a small discrepancy for a large value of $m$ from being misinterpreted as evidence in favor of normality merely because of the general convergence of $S_m^{\!(n)}(t)$ to $\psi_0(t)$. A more detailed motivation of this standardization including its relation to the asymptotic behavior of $S_m^{\!(n)}$ is provided in Section~S1 in the Supplementary Material.

To exploit information across different values of $m$, the test statistic of the SSTN is constructed based on the quantities $T_1^{\!(n)}, \dots, T_M^{\!(n)}$ rather than on a single standardized discrepancy. This is particularly important since, depending on the underlying distribution of the data, deviations from the theoretical self-similarity structure may be most pronounced at different values of $m$. While most distributions show a (much) slower convergence of $S_m^{\!(n)}(t)$ to $\psi_0(t)$ than the normal distribution, there also exist distributions such as the continuous uniform distribution for which this convergence is faster. Since the SSTN is based on the standardized discrepancies $T_m^{\!(n)}$, such differences in convergence speed affect how quickly consecutive empirical self-similarity transforms approach each other and thus influence the magnitude of these discrepancies. Therefore, considering a sufficiently large range of iteration levels $m=1,\dots,M$ allows the SSTN to capture self-similarity behavior across different scales. The choice of the maximum iteration level $M$ is discussed in Section~\ref{s:Numerical_Implementation}.

To, hence, ensure that the most pronounced deviations from the theoretical self-similarity structure are not diluted by averaging over iteration levels at which they are less pronounced, the standardized discrepancies are combined into the test statistic of the SSTN by taking the maximum of their absolute values so that this test statistic is given by
\begin{equation}
\label{f:test_statistic}
T^{\!(n)} := \max_{1\le m\le M} \big|T_m^{\!(n)}\big|.
\end{equation}

\subsection{Linearized representation of discrepancies} \label{s:linearized_representation_test_statistic}

A key step towards both an efficiently implementable test statistic for the SSTN and a tractable asymptotic analysis is to replace the discrepancies~\eqref{f:iterative_discrepancies} by a linearized representation of them that exposes their leading asymptotic structure since these discrepancies $D_m^{\!(n)}$ involve powers of standardized empirical characteristic functions, which are difficult to analyze directly. At the same time, this representation should allow, in practice, a numerically stable and computationally efficient evaluation of the discrepancy measure~\eqref{f:self-similarity_discrepancy_measures}. 

To this end, we first note that, under the null hypothesis that the underlying distribution $P_X$ is normal, the standardized empirical characteristic function \eqref{f:standardized_ecf} satisfies a functional central limit theorem, which allows us to express its deviation from the true characteristic function $\psi_0$ of the standard normal distribution in terms of a Gaussian limit. This is formalized in Proposition~\ref{prop:Convergence_of_Un}.

\begin{proposition}\label{prop:Convergence_of_Un}
Let $X_1, \dots, X_n$ be independent random variables that follow a $\mathcal N(\mu,\sigma^2)$ distribution, and let $\widehat\psi_X^*(t)$ be the standardized empirical characteristic function \eqref{f:standardized_ecf}. Furthermore, let $\psi_0$ and $\psi$ denote the characteristic functions of the standard normal and the $\mathcal N(\mu,\sigma^2)$ distribution, respectively, and define the process 
\begin{equation}\label{f:U_n}
U_n(t)
:=
\sqrt{n}\,\big(\widehat{\psi}_X^*(t) - \psi_0(t)\big),
\qquad t\in[-T,T].
\end{equation}
Then, for every fixed $T<\infty$, it holds that
\[
U_n
\xrightarrow{\mathrm{d}}
G_0
\qquad\text{in } C([-T,T];\mathbb{C}),
\]
where $G_0$ is a centered complex Gaussian process given by
\[
G_0(t)
= \exp\!\left(-it\,\frac{\mu}{\sigma}\right) 
  G\!\left(\frac{t}{\sigma}\right)
   - \frac{it}{\sigma}\,\psi_0(t)\,Y_1
   + \frac{t^2}{\sigma}\,\psi_0(t)\,Y_2.
\]
Here, $G$ is the centered complex Gaussian process arising as the weak limit
\[
\sqrt{n}\big(\widehat\psi_X - \psi\big)
\xrightarrow{\mathrm{d}}
G 
\qquad\text{in } C([-T/\sigma,T/\sigma];\mathbb{C})
\]
with covariance function
\[
\mathrm{Cov}\big(G(s),G(t)^{\mathrm c}\big)
= \psi(s-t) - \psi(s)\psi(-t),
\]
where $G(t)^{\mathrm c}$ denotes the complex conjugate of $G(t)$. Furthermore, $Y_1$ and $Y_2$ are uncorrelated normal random variables with mean $0$ and variance $\mathrm{Var}(Y_1)=\sigma^2$ and $\mathrm{Var}(Y_2)=\sigma^2/2$, which arise as the joint weak limit
\[
\sqrt{n}\begin{pmatrix} \bar X - \mu \\ S_X - \sigma \end{pmatrix}
\xrightarrow{\mathrm{d}}
\begin{pmatrix} Y_1 \\ Y_2 \end{pmatrix}.
\]
Moreover, the convergence statements above hold jointly, i.e.
\[
\Big(\sqrt{n}(\widehat\psi_X-\psi),\ \sqrt{n}(\bar X-\mu),\ \sqrt{n}(S_X-\sigma)\Big)
\ \xrightarrow{\mathrm d} \
(G,\ Y_1,\ Y_2)
\]
in $C([-T/\sigma,T/\sigma];\mathbb{C})\times\mathbb{R}^2$.
\end{proposition}

The proofs of the results in this section are provided in Section~S3 of the Supplementary Material.

This proposition establishes the weak convergence of the process~\eqref{f:U_n} in the function space $C([-T,T];\mathbb{C})$, i.e.\ convergence in distribution of the entire path $U_n$ toward a Gaussian limit process with respect to the supremum norm
\[
\|f\|_{\infty} := \sup_{t\in[-T,T]} |f(t)|.
\]

Building upon the result of Proposition~\ref{prop:Convergence_of_Un}, Lemma~\ref{lem:convergence_rate_D} provides an asymptotic expansion of the iterative discrepancies $D_m^{\!(n)}$ in terms of linear functionals of $U_n$.

\begin{lemma}\label{lem:convergence_rate_D}
Let $X_1, \dots, X_n$ be independent random variables that follow a $\mathcal N(\mu,\sigma^2)$ distribution and $D_m^{\!(n)}$ be the iterative discrepancies defined in \eqref{f:iterative_discrepancies}. Furthermore, define the transformed deviation
\begin{equation}\label{f:H_m^(n)}
H_m^{\!(n)}(t)
:=
(m+1)\,
\frac{U_n\!\big(t/\sqrt{m+1}\big)}{\psi_0\!\big(t/\sqrt{m+1}\big)}
-
m\,
\frac{U_n\!\big(t/\sqrt{m}\big)}{\psi_0\!\big(t/\sqrt{m}\big)},
\qquad t\in[-T,T].
\end{equation} 
Then, for each $m\in\{1,\dots,M\}$, there exists a term 
$r_m^{\!(n)} \in C([-T,T];\mathbb{C})$ such that
\[
D_m^{\!(n)}(t)
=
\frac{\psi_0(t)}{\sqrt{n}}H_m^{\!(n)}(t)
+ r_m^{\!(n)}(t),
\qquad t\in[-T,T],
\]
where
\[
\|r_m^{\!(n)}\|_{\infty,[-T,T]} = o_{\mathbb P}\big(n^{-1/2}\big),
\]
and for any $p \ge 1$, the family
\[
\big\{\|r_m^{\!(n)}\|_{\infty,[-T,T]}^p : n\in\mathbb N\big\}
\]
is uniformly integrable. Moreover, it holds that
\[
\|D_m^{\!(n)}\|_{\infty,[-T,T]} = O_{\mathbb P}\big(n^{-1/2}\big).
\]
\end{lemma}

This lemma shows that $D_m^{\!(n)}$ can be decomposed into a principal term of order $n^{-1/2}$ which involves the transformed deviation~\eqref{f:H_m^(n)} and a negligible rest term. This decomposition constitutes the essential analytical step in the asymptotic analysis of the test statistic, as it replaces a nonlinear functional of the empirical characteristic function by a linear expression with an explicit representation that is amenable to further analysis.

To derive an asymptotic representation of the discrepancy measures $\Delta_m^{\!(n)}$ based on the linearization in Lemma~\ref{lem:convergence_rate_D}, uniform moment control for the principal term needs to be ensured. Lemma~\ref{lem:moment_bound_phiH} provides the required bounds for the transformed deviations $H_m^{\!(n)}$ and guarantees uniform integrability.

\begin{lemma}\label{lem:moment_bound_phiH}
Let $X_1, \dots, X_n$ be independent random variables that follow a $\mathcal N(\mu,\sigma^2)$ distribution. Define the continuous functional
\[
\mathcal{T}: C([-T,T];\mathbb{C}) \to [0,\infty) 
\quad \text{with }
\mathcal{T}(f) := \int_{-T}^{T} w(t)\,\psi_0(t)^2\,|f(t)|^2\,dt,
\]
where $w$ is a bounded and nonnegative weighting function and $\psi_0$ is the characteristic function of the standard normal distribution. Moreover, define $U_n(t)$ as in \eqref{f:U_n} and $H_m^{\!(n)}$ as in \eqref{f:H_m^(n)}. Then, for any $q>0$,
\[
\sup_{n\in\mathbb N} \sup_{|t|\le T}\ \mathbb E\big[|U_n(t)|^q\big] < \infty.
\]
Moreover, for any $p \ge 1$ and each fixed $m\in\{1,\dots,M\}$,
\[
\sup_{n\in\mathbb N}
\mathbb E\!\left[\mathcal{T}\big(H_m^{\!(n)}\big)^p\right] < \infty.
\]
In particular, for any $p \ge 1$, the family $\{\mathcal{T}\big(H_m^{\!(n)}\big)^p : n \in \mathbb{N}\}$ is uniformly integrable.
\end{lemma}

The following theorem builds on the linear expansion of $D_m^{\!(n)}$ established in Lemma~\ref{lem:convergence_rate_D} together with the uniform moment bounds provided in Lemma~\ref{lem:moment_bound_phiH}.

\begin{theorem}\label{thm:moment_bound_nDelta}
Let $X_1, \dots, X_n$ be independent random variables that follow a $\mathcal N(\mu,\sigma^2)$ distribution, and let $\Delta_m^{\!(n)}$ be the self-similarity discrepancy measures defined in \eqref{f:self-similarity_discrepancy_measures}. Then, for each fixed $m\in\{1,\dots,M\}$, $\Delta_m^{\!(n)}$ admits the asymptotic expansion
\[
\Delta_m^{\!(n)}
=
\frac{1}{n} \int_{-T}^{T} w(t)\,\psi_0(t)^2\, \big|H_m^{\!(n)}(t)\big|^2\, dt
+ o_{\mathbb P}\big(n^{-1}\big),
\]
and consequently $\Delta_m^{\!(n)} = O_{\mathbb P}(n^{-1})$. Moreover, for any $p\ge 1$,
\[
\sup_{n\in\mathbb N} \mathbb E\big[\big(n\Delta_m^{\!(n)}\big)^p\big] < \infty.
\]
In particular, for any $p \ge 1$, the family $\{\big(n\Delta_m^{\!(n)}\big)^p : n \in \mathbb{N}\}$ is uniformly integrable.
\end{theorem}

The asymptotic representation of $\Delta_m^{\!(n)}$ obtained in Theorem~\ref{thm:moment_bound_nDelta} has an immediate practical implication for the determination of the test statistic of the SSTN. Since $\Delta_m^{\!(n)}$ is asymptotically equivalent to a quadratic functional of the transformed deviations $H_m^{\!(n)}$, the test statistic \eqref{f:test_statistic} can be determined numerically by computing the linearized discrepancy measures
\begin{equation}\label{f:linearized_discrepancy_measures}
Q_m^{\!(n)}
:= 
\int_{-T}^{T} 
w(t)\,\psi_0(t)^2 \, \big| H_m^{\!(n)}(t) \big|^2 \, dt,
\qquad m=1,\dots,M,
\end{equation}
which provide a numerically stable approximation for $n\Delta_m^{\!(n)}$. 

Theorem~\ref{thm:moment_bound_nDelta} shows that the discrepancy measures $\Delta_m^{\!(n)}$ admit a linearized approximation in terms of $n^{-1}Q_m^{\!(n)}$, with an asymptotically negligible deviation. This suggests that, for the purpose of determining the test statistic of the SSTN, the quantities $Q_m^{\!(n)}$ can be used in place of $\Delta_m^{\!(n)}$.

Accordingly, instead of standardizing the original discrepancies $\Delta_m^{\!(n)}$, we consider the standardized form
\begin{equation}
\label{f:standardized_discrepancies_tilde}
\widetilde T_m^{\!(n)} 
:= \frac{Q_m^{\!(n)} - \mathbb E_{H_0}\bigl[Q_m^{\!(n)}\bigr]}{\sqrt{\mathrm{Var}_{H_0}\bigl(Q_m^{\!(n)}\bigr)}},
\qquad m=1,\dots,M,
\end{equation}
and obtain the test statistic of the SSTN by
\begin{equation}
\label{f:test_statistic_tilde}
\widetilde T^{\!(n)} := \max_{1\le m\le M} \big|\widetilde T_m^{\!(n)}\big|.
\end{equation}

Lemma~\ref{lem:moment_bound_phiH} ensures that, under the null hypothesis, both $\mathbb E_{H_0}[Q_m^{\!(n)}]$ and $\mathrm{Var}_{H_0}(Q_m^{\!(n)})$ are finite for all $n \in \mathbb{N}$, while Theorem~\ref{thm:moment_bound_nDelta} yields the corresponding finiteness results for $n\Delta_m^{\!(n)}$. Hence, the standardized discrepancies in \eqref{f:standardized_discrepancies} and \eqref{f:standardized_discrepancies_tilde} are well-defined.

The linearized determination of $T^{\!(n)}$ provides a stable and computationally efficient basis for the application of the SSTN in practice, since the computation of $\widetilde T^{\!(n)}$ via the linearized discrepancies $Q_m^{\!(n)}$ avoids the repeated exponentiation of the standardized empirical characteristic function. Therefore, the linearized statistic $\widetilde T^{\!(n)}$, which is shown in the next section to be asymptotically equivalent to $T^{\!(n)}$, is used to approximate the null distribution of the SSTN via Monte Carlo simulation when the sample size $n$ is small or moderate, i.e.\ when $n < 100$. For larger sample sizes, the SSTN relies on the asymptotic null distribution derived there.

\subsection{Asymptotic null distribution of the test statistic}\label{s:asymptotic_test_statistic}

To derive the asymptotic null distribution of the test statistic of the SSTN, we first introduce the corresponding limit quantities that arise from the linearized representation of the discrepancies obtained in Theorem~\ref{thm:moment_bound_nDelta}. For each $m\in\{1,\dots,M\}$, define the limit variable
\begin{equation}\label{f:Q_m}
Q_m := \int_{-T}^{T} w(t)\,\psi_0(t)^2\,\big|H_m(t)\big|^2\,dt
\end{equation}
with
\[
H_m(t)
:=
(m+1)\,\frac{G_0\bigl(t/\sqrt{m+1}\bigr)}{\psi_0\bigl(t/\sqrt{m+1}\bigr)}
-
m\,\frac{G_0\bigl(t/\sqrt{m}\bigr)}{\psi_0\bigl(t/\sqrt{m}\bigr)},
\]
where $G_0$ is the Gaussian limit process introduced in Proposition~\ref{prop:Convergence_of_Un}. 

The variable $Q_m$ arises as a quadratic functional of the Gaussian limit process $G_0$. Therefore, this variable provides the weak limit of the rescaled discrepancy measure $n\Delta_m^{(n)}$ under the null hypothesis, as Theorem~\ref{thm:asymptotic_behavior_delta} shows.

\begin{theorem}\label{thm:asymptotic_behavior_delta}
Let $X_1, \dots, X_n$ be independent random variables that follow a
$\mathcal N(\mu,\sigma^2)$ distribution and $\Delta_m^{\!(n)}$,
$m = 1, \dots, M$, be the self-similarity discrepancy measures defined in
\eqref{f:self-similarity_discrepancy_measures}. Then
\[
n\big(\Delta_1^{\!(n)},\dots,\Delta_M^{\!(n)}\big)
\ \xrightarrow{\mathrm d} \
\big(Q_1,\dots,Q_M\big)
\quad\text{as }n\to\infty,
\] 
where $Q_1,\dots,Q_M$ are the limit variables defined in \eqref{f:Q_m}.
\end{theorem}

As in Section~\ref{s:linearized_representation_test_statistic}, the proofs of the results in this section can be found Section~S3 of the Supplementary Material.

To make the limiting distribution in Theorem~\ref{thm:asymptotic_behavior_delta} applicable, we must verify that the limits $Q_m$, $m = 1, \dots, M$, possess finite moments and that the expectations and variances of $n\Delta_m^{\!(n)}$ converge to those of $Q_m$. The following lemma provides these moment convergence results.

\begin{lemma}\label{lem:mean_var_expansion}
Let $X_1, \dots, X_n$ be independent random variables with $\mathcal N(\mu,\sigma^2)$ distribution and $m\in\{1,\dots,M\}$ be fixed. Then, for any $p \ge 1$, the limit $Q_m$ defined in \eqref{f:Q_m} satisfies $\mathbb E[Q_m^p]<\infty$ and
\[
\mathbb E\!\left[\bigl(n\Delta_m^{\!(n)}\bigr)^p\right] \rightarrow \mathbb E\big[Q_m^p\big]
\qquad \text{as } n \to \infty.
\]
Consequently, for the finite constants
\[
\mu_m := \mathbb E[Q_m]
\quad \text{and} \quad
\sigma_m^2 := \mathrm{Var}(Q_m)
\]
with $\sigma_m^2 > 0$ it holds that
\[
\mathbb E\big[n\Delta_m^{\!(n)}\big] \rightarrow \mu_m
\quad \text{and} \quad
\mathrm{Var}\big(n\Delta_m^{\!(n)}\big) \rightarrow \sigma_m^2
\qquad \text{as } n\to\infty.
\]
\end{lemma}

Combining the results of Lemma~\ref{lem:mean_var_expansion} with those of Theorem~\ref{thm:moment_bound_nDelta}, we obtain the following asymptotic equivalence between the original test statistic $T^{\!(n)}$ of the SSTN given by~\eqref{f:test_statistic} and its linearized counterpart $\widetilde T^{\!(n)}$ given by~\eqref{f:test_statistic_tilde}.

\begin{proposition}\label{prop:convergence_tildeT-T} 
Let $T_m^{\!(n)}$, $m = 1, \dots, M$, be the standardized discrepancies defined in \eqref{f:standardized_discrepancies} and $\widetilde T_m^{\!(n)}$ be the linearized standardized discrepancies defined in \eqref{f:standardized_discrepancies_tilde}. Then it holds that
\[
\max_{1\le m\le M}\big|\widetilde T_m^{\!(n)}-T_m^{\!(n)}\big|
\xrightarrow{\mathbb P} 0.
\]
In particular, we have
\[
\big|\widetilde T^{\!(n)}-T^{\!(n)}\big|
\xrightarrow{\mathbb P} 0.
\]
\end{proposition}

Based on the results of Lemma~\ref{lem:mean_var_expansion}, the asymptotic standardized discrepancies can be defined as 
\[
T_m := \frac{Q_m - \mu_m}{\sigma_m}, \qquad m=1,\dots,M,
\]
leading to the limit statistic
\begin{equation}\label{f:asymptotic_T}
T := \max_{1\le m\le M} \big|T_m\big|.    
\end{equation}
As Theorem~\ref{thm:asymptotic_distribution_SSTN} shows, these random variables represent the asymptotic counterparts of the standardized discrepancies $T_m^{\!(n)}$ and the SSTN test statistic $T^{\!(n)}$. The following theorem confirms this correspondence.

\begin{theorem}\label{thm:asymptotic_distribution_SSTN}
Let $X_1, \dots, X_n$ be independent random variables that follow a $\mathcal N(\mu,\sigma^2)$ distribution and let $T_m^{\!(n)}$, $m = 1, \dots, M$, be the standardized discrepancies defined in \eqref{f:standardized_discrepancies}. Then, it holds that
\[
\left(T_1^{\!(n)},\dots,T_M^{\!(n)}\right) \
\xrightarrow{\mathrm d} \
\big(T_1,\dots,T_M\big) \quad \text{as }n\to\infty.
\]
Consequently, for the overall test statistic \eqref{f:test_statistic} of the SSTN, we have
\[
T^{\!(n)} \ \xrightarrow{\mathrm d} \ T.
\]
Moreover, for the linearized form ~\eqref{f:test_statistic_tilde}, it holds that
\[
\widetilde T^{\!(n)} \ \xrightarrow{\mathrm d} \ T.
\]
\end{theorem}

The asymptotic null distribution of the test statistic~\eqref{f:asymptotic_T} of the SSTN cannot be specified in closed form, and therefore, needs to be approximated by, e.g., a Monte Carlo simulation. A procedure for this approximation is discussed in the following section.

\section{Performing the SSTN}
\label{s:Numerical_Implementation}

This section describes the numerical procedures for applying the SSTN and the steps required to make inference. As discussed in Section~\ref{s:asymptotic_test_statistic}, the asymptotic null distribution based on the test statistic~\eqref{f:asymptotic_T} is employed when data for $n \ge 100$ observations are available. In the following section, we discuss how this asymptotic null distribution can be obtained. We then describe how the sample-size-specific null distribution can be determined. Finally, we present the test decision procedure of the SSTN.

\subsection{Approximation of the asymptotic null distribution}
\label{s:simulate_T0}

Since $Q_m$ is fully characterized by the Gaussian limit process $G_0$ introduced in Proposition~\ref{prop:Convergence_of_Un}, realizations of $G_0$ are evaluated in the SSTN on a finite, equidistant, and symmetric grid $t_1,\dots,t_H \in [-T,T]$, where
\begin{equation}\label{f:t_h}
t_h
=
-T + \frac{2(h-1)}{H-1}T,
\qquad h = 1,\dots,H,
\end{equation}
yielding the vector $\big(G_0(t_1),\dots,G_0(t_H)\big)$ and thereby realizations of the limit variable $Q_m$.

For this purpose, the joint limit $(G, Y_1, Y_2)$ from Proposition~\ref{prop:Convergence_of_Un}, which defines the Gaussian limit process $G_0$ by
\[
G_0(t)
= \exp\!\left(-it\,\frac{\mu}{\sigma}\right) 
  G\!\left(\frac{t}{\sigma}\right)
   - \frac{it}{\sigma}\,\psi_0(t)\,Y_1
   + \frac{t^2}{\sigma}\,\psi_0(t)\,Y_2,
\]
is simulated and evaluated on this grid. Writing the complex-valued Gaussian process $G$ as
\[
G(t) = G_1(t) + i\,G_2(t),
\]
with real-valued Gaussian processes $G_1$ and $G_2$, the joint limt can be represented by a real-valued and centered $(2H+2)$-dimensional Gaussian vector
\[
\mathbf G
:=
\big(G_1(t_1),\dots,G_1(t_H),\;
     G_2(t_1),\dots,G_2(t_H),\;
     Y_1,\,Y_2\big).
\]
An explicit characterization of the joint covariance structure of this vector $\mathbf G$ is derived in Theorem~S1 in Section~S2 of the Supplementary Material, which thus enables the simulation of $\mathbf G$. This simulation is performed under the standard normal distribution so that the domain $[-T/\sigma,T/\sigma]$ considered in Theorem~S1 is given by $[-T,T]$.

Specifically, each Monte Carlo replication under the standard normal model is obtained by simulating the random vector $\mathbf G$ with covariance matrix $\boldsymbol{\Sigma}$ given in Theorem~S1, which simplifies in this case to the block matrix
\[
\boldsymbol{\Sigma}
=
\begin{pmatrix}
\boldsymbol{\Sigma}_{11} & \mathbf 0_{H \times H} & \mathbf 0_H & c_{1,Y_2} \\[1ex]
\mathbf 0_{H \times H} & \boldsymbol{\Sigma}_{22} & c_{2,Y_1} & \mathbf 0_H \\[1ex]
\mathbf 0_H^{\!\top} & c_{2,Y_1}^{\!\top} & 1 & 0 \\[1ex]
c_{1,Y_2}^{\!\top} & \mathbf 0_H^{\!\top} & 0 & \tfrac12
\end{pmatrix},
\]
where $\boldsymbol{\Sigma}_{11}$ and $\boldsymbol{\Sigma}_{22}$ are the $H\times H$ covariance matrices with entries
\begin{align*}
(\boldsymbol{\Sigma}_{11})_{hl}
&=\mathrm{Cov}\big(G_1(t_h),G_1(t_l)\big)
=\frac12\Big(\psi_0(t_h-t_l)+\psi_0(t_h+t_l)-2\psi_0(t_h)\psi_0(t_l)\Big),\\
(\boldsymbol{\Sigma}_{22})_{hl}
&=\mathrm{Cov}\big(G_2(t_h),G_2(t_l)\big)
=\frac12\Big(\psi_0(t_h-t_l)-\psi_0(t_h+t_l)\Big)
\end{align*}
for $h,l=1,\dots,H$, and the cross-covariance vectors are given by
\[
c_{2,Y_1}=\bigl(t_1\psi_0(t_1),\dots,t_H\psi_0(t_H)\bigr)^\top
\quad \text{and} \quad
c_{1,Y_2}=\bigl(-\tfrac12 t_1^2\psi_0(t_1),\dots,-\tfrac12 t_H^2\psi_0(t_H)\bigr)^\top.
\]

The resulting realizations of $\mathbf G$ are then combined pointwise to construct the discretized limit process
\[
G_0(t_h)
=
G(t_h)
- i\,t_h\,\psi_0(t_h)\,Y_1
+ t_h^2\,\psi_0(t_h)\,Y_2,
\qquad h=1,\dots,H.
\]

Based on these discretized values, a continuous approximation $\widetilde G_0$ on $[-T,T]$ is formed by linear interpolation. The transformed deviations $H_m$ are then estimated on this grid by
\[
H_m(t_h)
\approx
(m+1)\,\frac{\widetilde G_0\bigl(t_h/\sqrt{m+1}\bigr)}{\psi_0\bigl(t_h/\sqrt{m+1}\bigr)}
-
m\,\frac{\widetilde G_0\bigl(t_h/\sqrt{m}\bigr)}{\psi_0\bigl(t_h/\sqrt{m}\bigr)},
\qquad h=1,\dots,H,\ \ m=1,\dots,M.
\]

Finally, the limiting discrepancy functionals 
\[
Q_m
\approx
\sum_{h=1}^H
\exp(-\beta t^2)\,\psi_0(t_h)^2\,|H_m(t_h)|^2\,\Delta t,
\qquad m=1,\dots,M,
\]
are approximated by numerical integration using Riemann sums, where $\beta>0$ and $\Delta t=2T/(H-1)$.

In the SSTN, this procedure is repeated $B = 10{,}000$ times to estimate the joint distribution of $(Q_1,\dots,Q_M)$ under the null hypothesis of normality. By default, the maximum iteration level is set to $M=20$, the grid size to $H=100$, and the upper integration bound to $T=4$. While, in principle, increasing the number of iterations $M$ may provide a finer assessment of self-similarity, and increasing $H$ and $T$ may marginally improve the accuracy in the numerical approximation of characteristic functions, the chosen parameter values were found to be sufficiently large for a stable performance of the SSTN, as further increases did not lead to noticeable changes in empirical performance.

For the choice of the parameter $\beta$ of the weighting function $w(t)=\exp(-\beta t^2)$, it is crucial to note that this function assigns larger weights to the center of the Fourier transform in order to reduce the influence of errors in the more susceptible tails. This effect increases with larger values of $\beta$. However, choosing $\beta$ too large may lead to a loss of information so that a balanced choice is required. In a simulation study considering different values for this parameter, $\beta = 2$ turned out to be a suitable choice in the SSTN (see Section~S4 of the Supplementary Material).

To obtain a stable standardization of the limiting variables,
\[
\mu_m^\text{asy} = \mathbb E_{H_0}[Q_m]
\quad \text{and} \quad
\sigma_m^\text{asy} = \sqrt{\mathrm{Var}_{H_0}(Q_m)}
\]
are estimated in an independent Monte Carlo simulation step by generating a second set of $B$ independent realizations of the limiting variables $Q_m$ under the same setting as in the original sampling. The arithmetic mean $\widehat\mu_m^\text{asy}$ and the empirical standard deviation $\widehat\sigma_m^\text{asy}$ are then computed from these samples to estimate $\mu_m^\text{asy}$ and $\sigma_m^\text{asy}$.

The empirical distribution of the limit statistic~\eqref{f:asymptotic_T} is subsequently obtained by determining
\[
t_{\mathrm{asy}}^{\!(b)}
=
\max_{1\le m\le M}
\left|
\frac{q_m^{\!(b)} - \widehat\mu_m^\text{asy}}{\widehat\sigma_m^\text{asy}}
\right|,
\qquad b=1,\dots,B,
\]
based on the realizations $\bigl(q_1^{\!(b)},\dots,q_M^{\!(b)}\bigr)$ of $(Q_1,\dots,Q_M)$.

\subsection{Approximation of the finite-sample null distribution}\label{s:finite_sample_simulation}

For smaller sample sizes ($n < 100$), the SSTN relies on a sample size-specific Monte Carlo approximation of the null distribution, which is obtained by repeated computation of the linearized test statistic \eqref{f:test_statistic_tilde} under independent samples from the standard normal distribution. 

To this end, $B = 10{,}000$ independent samples 
\[
\mathbf x^{\!(b)} = \Bigl(x_1^{\!(b)},\dots,x_n^{\!(b)}\Bigr), \quad b=1,\dots,B,
\]
are generated, each consisting of $n$ observations drawn from a standard normal distribution. For each sample $\mathbf x^{\!(b)}$, $b = 1, \dots, B$, the standardized empirical characteristic function
\[
\widehat\psi_{\mathbf x^{\!(b)}}^*(t)
=
\exp\!\left(-it\,\frac{\bar x^{\!(b)}}{s_X^{\!(b)}}\right)
\cdot 
\frac{1}{n}\sum_{k=1}^n 
\exp\!\left( it\,\frac{x_k^{\!(b)}}{s_X^{\!(b)}} \right)
\]
is computed, where $\bar x^{\!(b)}$ denotes the arithmetic mean and $s_X^{\!(b)}$ the empirical standard deviation of the samples $\mathbf x^{\!(b)}$. This function is then employed to determine the empirical deviation process
\[
u_n^{\!(b)}(t_h)
=
\sqrt{n}\Bigl(\widehat\psi_{\mathbf x^{\!(b)}}^*(t_h) - \psi_0(t_h)\Bigr), \quad h = 1, \dots, H,
\]
on the equidistant grid $t_1,\dots,t_H\in[-T,T]$ where, as in the determination of the asymptotic null distribution, $H = 100$ and $T = 4$ is used.

Based on $u_n^{\!(b)}(t_1), \dots, u_n^{\!(b)}(t_H)$, $b = 1, \dots, B$, the transformed deviations $h_{m,n}^{\!(b)}$, $m = 1, \dots, M$, are computed by 
\[
h_{m,n}^{\!(b)}(t)
=
(m+1)\,
\frac{\widetilde u_n^{\!(b)}\!\big(t/\sqrt{m+1}\big)}{\psi_0\!\big(t/\sqrt{m+1}\big)}
-
m\,
\frac{\widetilde u_n^{\!(b)}\!\big(t/\sqrt{m}\big)}{\psi_0\!\big(t/\sqrt{m}\big)},
\]
where $\widetilde u_n^{\!(b)}$ denotes the piecewise linear interpolation of $u_n^{\!(b)}$. Afterwards, the linearized discrepancy measures $q_{m,n}^{\!(b)}$, $m=1,\dots,M$, are computed as numerical approximations of \eqref{f:linearized_discrepancy_measures} via the Riemann sum
\[
q_{m,n}^{\!(b)} =
\sum_{h=1}^{H}
\exp(-\beta t_h^2)\,\psi_0(t_h)^2\, \big|h_{m,n}^{\!(b)}(t_h)\big|^2\,\Delta t,
\qquad m=1,\dots,M,
\]
where $\Delta t = 2T/(H-1)$. As discussed in Section~\ref{s:simulate_T0}, we typically set $\beta = 2$ and $M = 20$.

The resulting values $q_{m,n}^{\!(b)}$, $b=1,\dots,B$, are then used to estimate the sample size-specific mean and variance under the null hypothesis by
\[
\hat\mu_{m,n} := \frac{1}{B}\sum_{b=1}^B q_{m,n}^{\!(b)}
\quad \text{and} \quad
\hat\sigma_{m,n}^2 := 
\frac{1}{B-1}\sum_{b=1}^B\Bigl(q_{m,n}^{\!(b)}-\hat\mu_{m,n}\Bigr)^2.
\]

Subsequently, each Monte Carlo replicate is standardized according to
\[
\tilde t_{m,n}^{(b)} 
= \frac{q_{m,n}^{\!(b)} - \hat\mu_{m,n}}{\hat\sigma_{m,n}},
\qquad m=1,\dots,M,
\]
and the value of the test statistic for the $b$-th sample, $b=1,\dots,B$, is computed by
\[
\tilde t_n^{(b)} = 
\max_{1\le m\le M}\big|\tilde t_{m,n}^{(b)}\big|.
\]
The empirical distribution of $\tilde t_n^{(1)}, \dots, \tilde t_n^{(B)}$ is then used as an approximation of the finite-sample null distribution of $\tilde T_n$.

\subsection{Test decision}
\label{s:test_decision}

Given an observed sample vector $\mathbf x=(x_1,\dots,x_n)^\top$ that should be tested for normality, the test decision is made in the SSTN as follows.

Based on this sample vector $\mathbf{x}$, the linearized discrepancy measures $q_{m,n}^{\mathrm{obs}}$, $m=1,\dots,M$, are computed as described in Section~\ref{s:finite_sample_simulation}. For $n < 100$, the sample size-specific estimates $\hat\mu_{m,n}$ and $\hat\sigma_{m,n}$, and otherwise, the asymptotic estimates $\widehat\mu_m^{\mathrm{asy}}$ and $\widehat\sigma_m^{\mathrm{asy}}$ are used to compute the observed value
\[
t^{\mathrm{obs}}
=
\max_{1\le m\le M}
\left|
\frac{q_m^{\mathrm{obs}} - \widehat\mu_m}{\widehat\sigma_m}
\right|
\]
of the test statistic of the SSTN for the sample $\mathbf{x}$. The $p$-value of the SSTN is then computed by
\[
p
=
\frac{1}{B}
\sum_{b=1}^B
\mathbf 1\bigl(t^{(b)} \ge t^{\mathrm{obs}}\bigr),
\]
where $t^{(b)}$, $b = 1, \dots, B$, are the Monte Carlo replicates of this test statistic computed as described in Section~\ref{s:simulate_T0} or~\ref{s:finite_sample_simulation}, respectively, to estimate the null distribution.

\section{Simulation Study} \label{s:Simulation_study} 

To evaluate the performance of the SSTN, we conducted a simulation study that covers a broad range of scenarios for testing for normality. As noted in Section~\ref{s:Numerical_Implementation}, the choice $\beta = 2$ for the weighting function \eqref{f:weighting} turned out to be a promising specification and, therefore, we present the results based on this value in the current section. A more detailed investigation of the influence of $\beta$ on the performance of the SSTN is provided in Section~S4 of the Supplementary Material, in which we compare the empirical behavior of the SSTN for different values of $\beta$.

\subsection{Simulation setup}

In this simulation study, the SSTN was compared to five widely used normality tests proposed by Shapiro-Wilk \cite{shapiro1965}, Anderson-Darling \cite{anderson1954}, Jarque-Bera \cite{jarque1980}, Lilliefors \cite{lilliefors1967}, and D'Agostino-Pearson \cite{dagostino1973}. All tests were performed at the nominal significance level $\alpha = 0.05$. 

To ensure a comprehensive comparison, we considered multiple distribution families under varying sample sizes and distribution-specific parameters. In total, eight distributions, including the normal distribution to assess type~I error control as well as the Gamma distribution, $\chi^2$ distribution, lognormal distribution, Weibull distribution, Student’s $t$ distribution, mixture of two normal distributions, and a convolution of a uniform distribution with a normal distribution were examined. The distributions were chosen similarly to those in the study on the power comparison of normality tests in \cite{arnastauskaite2021}.

Each of these distributions were considered for $n \in \{10, 25, 50, \allowbreak 100, 250, 500\}$ and six parameter values specific to each distribution, resulting in $36$ configurations per distribution. The values of the parameters are summarized in Table~\ref{t:settings}. The parameter values in each simulation scenario were chosen to yield gradually decreasing effect sizes, ensuring that for the smallest parameter setting the power of most tests approaches 1 and for the largest parameter value the distribution most closely resembles the normal distribution. This design thus enables an informative comparison across uniformly spaced grids of parameter values. 

Note that the $t$ distribution with one degree of freedom, i.e.\ the standard Cauchy distribution, was included, representing a self-similar distribution without finite expectation or variance. Moreover, in one setting of the convolution of a uniform and a normal distribution, the variance parameter $\sigma^2$ of the normal distribution was set to zero, which reduces this case to the uniform distribution.

\begin{table}[t]
\centering
\caption{Simulation setups considered in the simulation study.}
\label{t:settings}
\renewcommand{\arraystretch}{1.15}
\begin{tabular}{l l l}
\hline
\textbf{Distribution} 
& \textbf{Parameter} 
& \textbf{Parameter values} \\
\hline
$\mathcal N(a-3,\,9/a^2)$ 
& $a$ 
& $\{1,2,3,4,5,6\}$ \\

$\Gamma(\theta,1)$ 
& $\theta$ 
& $\{1,4,6,8,10,12\}$ \\

$\chi^2_m$ 
& $m$ 
& $\{3,6,9,12,15,18\}$ \\

$\mathrm{Lognormal}\bigl(0,(7/6-d)^2\bigr)$ 
& $d$ 
& $\{1/6,2/6,3/6,4/6,5/6,1\}$ \\

$\mathrm{Weibull}(k,1)$ 
& $k$ 
& $\{0.5,1,1.5,2,2.5,3\}$ \\

$t_\nu$ 
& $\nu$ 
& $\{1,2,3,4,5,6\}$ \\

$0.6\,\mathcal N(-1,1) + 0.4\,\mathcal N(4-b,2)$
& $b$ 
& $\{0,1,2,3,4,5\}$ \\

$\mathcal U(-3,3) * \mathcal N(0,\sigma^2)$
& $\sigma$ 
& $\{0,0.2,0.4,0.6,0.8,1\}$ \\
\hline
\end{tabular}
\end{table}

For each of the 36 settings, $R = 1{,}000$ data sets were generated and the proportion of rejections of the null hypothesis for each test was calculated to assess its statistical power at the significance level $\alpha = 0.05$.

\subsection{Results of the simulation study}

Before comparing the power of the SSTN and the other tests, we first assessed their type~I error control at the level $\alpha = 0.05$, considering the settings with the normal distribution, i.e.\ the scenarios under the null hypothesis. To quantify this, we computed acceptance bands for the expected rejection proportion under the null hypothesis \cite{delicado2007} by considering a binomial distribution with $R = 1{,}000$ replications and a probability $\alpha = 0.05$. This led to a two-sided $95\%$ confidence band, reaching form $0.037$ to $0.064$.

\begin{figure}[t]
 \centerline{\includegraphics[width=5.75in]{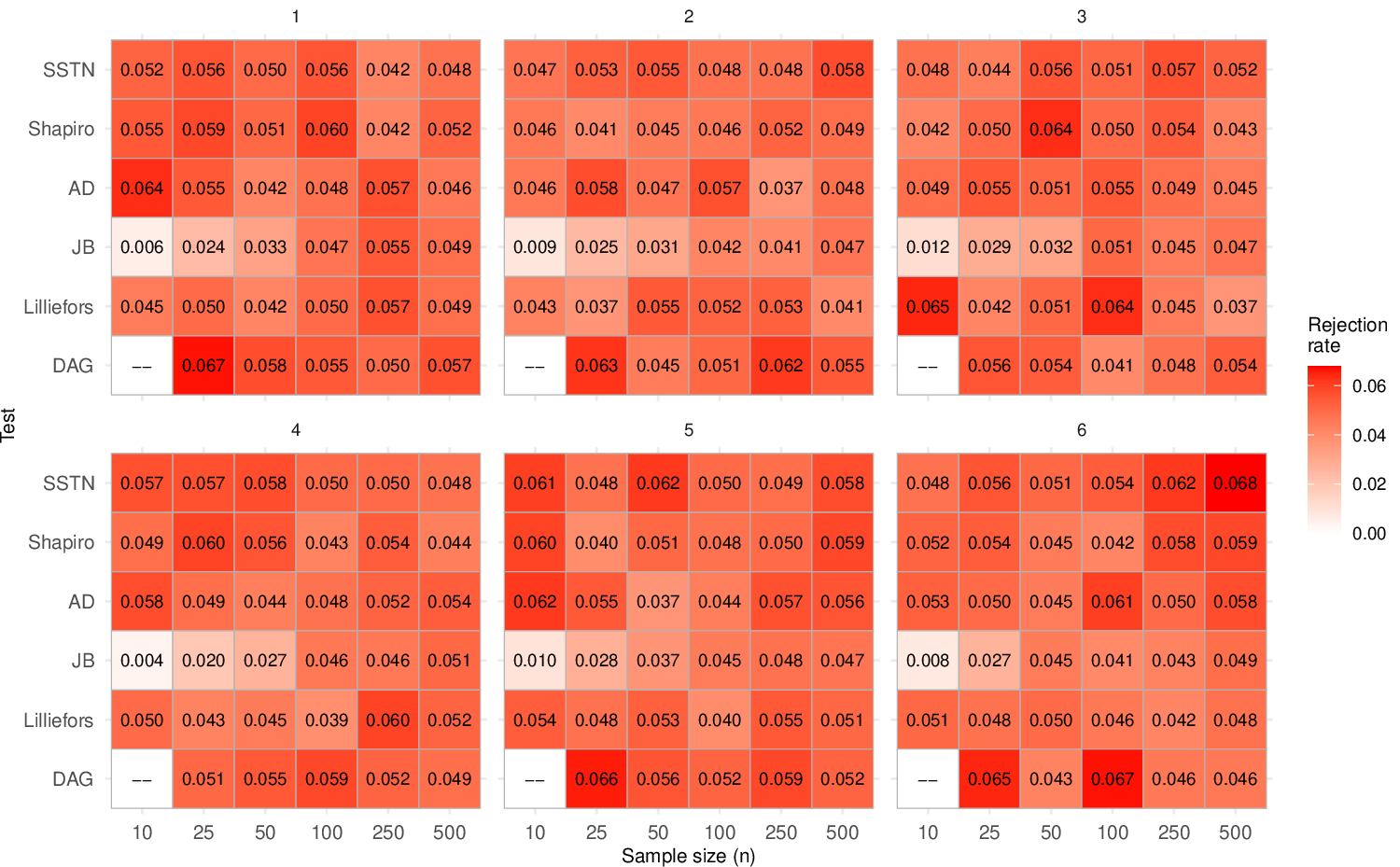}}
 \caption{Heatmaps of empirical rejection rates under the null hypothesis (normal distribution) across sample sizes and tests, separated by the underlying parameter values. Since the D’Agostino–Pearson test was developed for sample sizes $\geq 20$ it is not applicable for $n = 10$, so these cases are omitted.}
\label{fig:Normal}
\end{figure}

The results for the rejection rates under the null hypothesis are shown in Figure~\ref{fig:Normal}. For the SSTN, the rejection rates for all but one configuration lay within the acceptance band, indicating overall adequate type~I error control. The remaining configuration only slightly exceeded the upper bound (0.068 at $n = 500$ for a parameter value of 6). In comparison, the type~I error rates of the Shapiro–Wilk and Anderson-Darling tests always remained within the acceptance bands. Furthermore, the Lilliefors test showed one exceedance and the D’Agostino–Pearson test four exceedances. The rejection rates of the Jarque–Bera test, however, frequently fell below the lower bound, particularly for small sample sizes $n$, indicating an overly conservative behavior.

Under the alternative hypothesis, the power of all considered tests increased across all procedures with larger sample sizes and with distributional parameters that induce stronger deviations from normality. For asymmetric distributions, namely the Gamma, $\chi^2$, lognormal, and Weibull distributions, the SSTN outperformed all competitors except for the Shapiro–Wilk test in almost all settings.

\begin{figure}[H]
\vspace{1cm}
 \centerline{\includegraphics[width=5.75in]{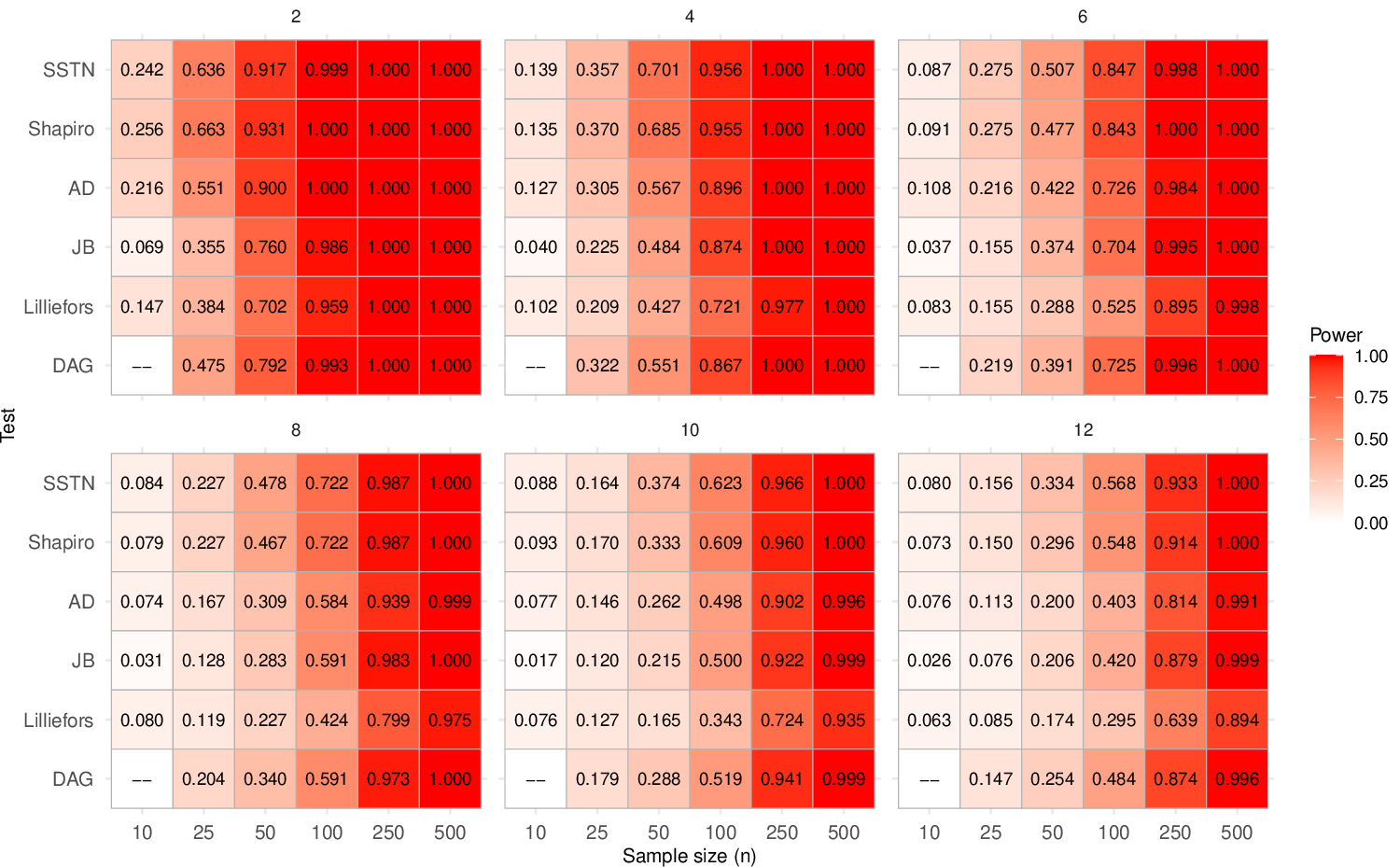}}
 \caption{Heatmaps of statistical power under the Gamma distribution, evaluated across sample sizes and tests and separated by the underlying parameter values. Cases with $n = 10$ for the D’Agostino–Pearson test are omitted.}
\label{fig:Gamma}
\end{figure}

Compared to the Shapiro–Wilk test, the SSTN showed in the case of the Gamma distribution just slightly weaker results for the first parameter value when $n < 100$, while it gained a small advantage for the larger parameter values (see Figure \ref{fig:Gamma}). All other figures referred to in this section can be found in the~\nameref{Appendix}. For the $\chi^2$ distribution, it can be seen from Figure \ref{fig:Chisq} that the pattern closely resembled these results, with both tests showing a very similar performance. Similar applies to the settings with a lognormal distribution (see Figure \ref{fig:Lognormal}). For the Weibull distribution, it is shown in Figure \ref{fig:Weibull} that the SSTN remained very competitive overall, but its performance tended to be marginally weaker than that of the procedure of Shapiro–Wilk.

For the symmetric and heavy-tailed $t$ distribution and for the mixture of normal distributions, the performance of the SSTN was among the best in specific parameter ranges. Specifically, for the Student’s $t$ distribution, the test performed best for $n \in \{10,25\}$ at two degrees of freedom and exhibited comparatively strong performance across all settings, with only slightly higher power values observed for the Jarque-Bera and D’Agostino-Pearson procedures (see Figure~\ref{fig:t}). For the mixture of normal distributions, the SSTN dominated for moderate to large distribution parameters, while it performed weaker for the parameter values $0$ and $1$ for which the procedures of Shapiro–Wilk, Anderson–Darling, and Lilliefors exhibited superior results (see Figure \ref{fig:Normal_mixture}).

The convolution of a uniform distribution with a normal distribution forms a special case due to the fast convergence associated with the uniform component (see Section~S1 in the Supplementary Material). The results of this case are depicted in Figure \ref{fig:Uniform}. Here, the approach of D’Agostino–Pearson achieved clearly the highest power for $n\ge 50$ across all tests considered. Even though the SSTN consistently outperformed the procedures of Jarque–Bera and Lilliefors for larger values of $n$ in this distribution case, its power often fell noticeably below those of the Shapiro–Wilk and D’Agostino–Pearson tests. The SSTN reached high power ($\ge 0.7$) in the pure uniform case for $n\ge 100$, whereas the procedures of Shapiro–Wilk and D’Agostino–Pearson did so already for $n=50$.

\section{Conclusions}\label{s:Conclusions}

In this article, we introduced the Self-Similarity Test for Normality (SSTN), a novel testing procedure for assessing the normality of observed data based on the self-similarity property of the normal distribution. For this, the SSTN assesses departures from this structure by operating on the level of characteristic functions. Specifically, the method constructs a deviation process for the standardized empirical characteristic function and quantifies discrepancies from self-similarity across multiple scales, thereby assessing departures from normality. These deviations are combined into a single max-type statistic, and inference is based on a pre-simulated Monte Carlo approximation under the null hypothesis.

A comprehensive simulation study covering a wide range of distributions with varying parameters and sample sizes demonstrated that the SSTN was at least competitive with five well-established normality tests in almost all scenarios. In several cases, particularly for samples drawn from asymmetric distributions, the SSTN often even showed the strongest performance. Only in very few scenarios, most notably under the uniform distribution, the SSTN exhibited weaker results.

The SSTN utilizes the weighted squared $L^2$-distance to quantify deviations from self-similarity by measuring discrepancies between successive self-similarity transforms of the standardized empirical characteristic function. While this weighted distance measure is motivated by its property that it reduces the influence of the sensitive tails of the estimated characteristic function in order to stabilize the comparison, other weighting schemes or distance measures are conceivable. For instance, data-driven or covariance-based weighting schemes adapting to the variability of the empirical characteristic function across the domain could be considered. It would, therefore, be of interest for future research to investigate how the SSTN performs when employing such weighting schemes. Moreover, it might be interesting to extend the SSTN to multivariate settings.

The SSTN is implemented in the R package \textbf{sstn}, which is freely available on CRAN at \textcolor{blue}{https://cran.r-project.org/web/packages/sstn}.

\section*{Acknowledgement}

We thank Arnak Dalalyan for insightful and constructive comments that substantially improved the manuscript.

This work has been supported by the Research Training Group “Biostatistical Methods for High-Dimensional Data in Toxicology” (RTG 2624, Project P4) funded by the Deutsche Forschungsgemeinschaft (DFG, German Research Foundation – Project Number 427806116).

\section*{Supplementary Material}

The Supplementary Material provides additional technical statements that support the methodological development in the main text, contains the proofs of all results, and includes a further simulation study illustrating and evaluating the effect of different choices of the weighting parameter in the SSTN.

\newpage

\begin{appendix}

\section*{Appendix}\label{Appendix}

\setcounter{figure}{0}
\renewcommand{\thefigure}{A\arabic{figure}}

\begin{figure}[H]
 \centerline{\includegraphics[width=5.75in]{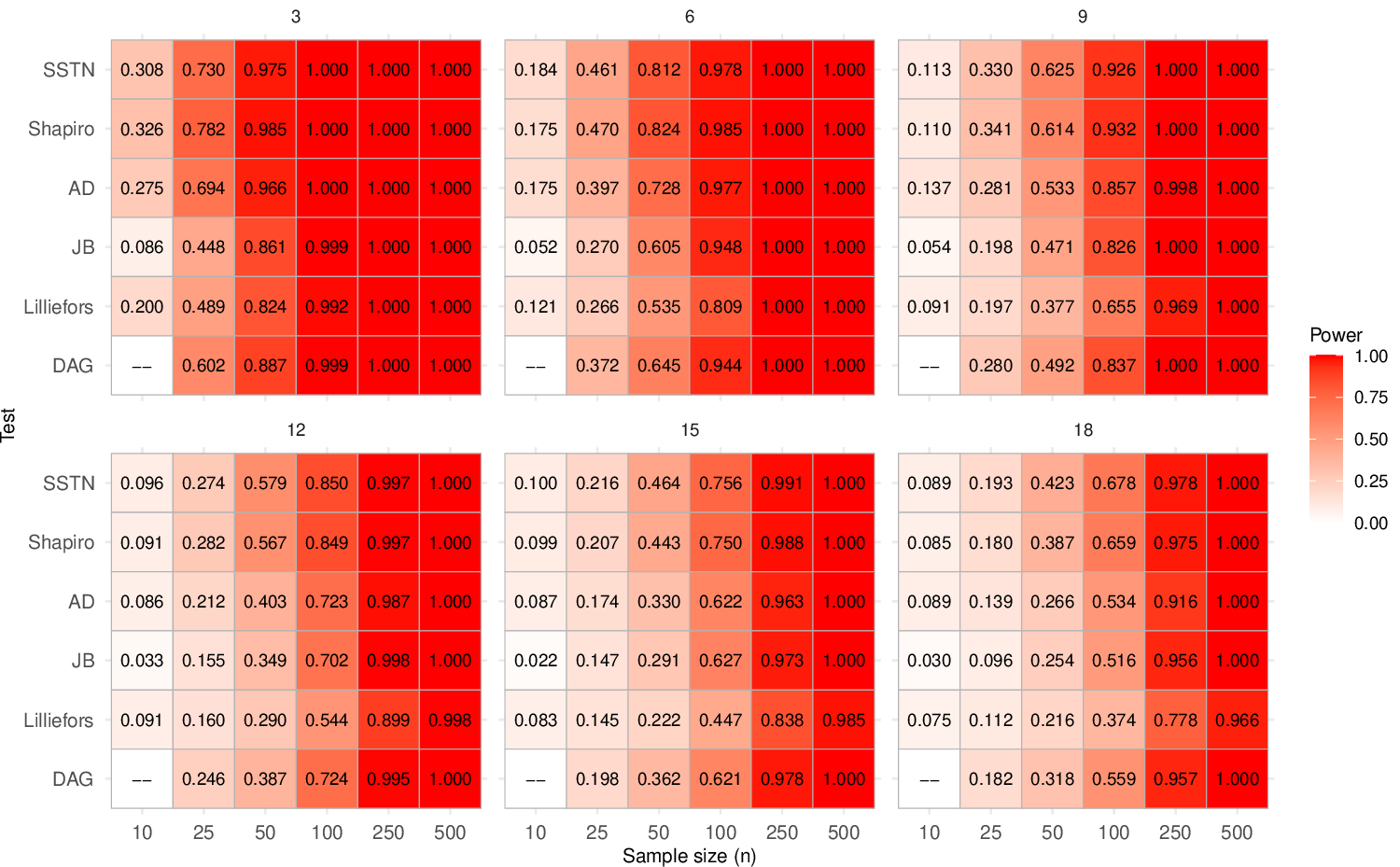}}
 \caption{Heatmaps of statistical power under the $\chi^2$ distribution, evaluated across sample sizes and tests and separated by the underlying parameter values. Cases with $n = 10$ for the D’Agostino–Pearson test are omitted.}
\label{fig:Chisq}
\end{figure}

\begin{figure}[H]
 \centerline{\includegraphics[width=5.75in]{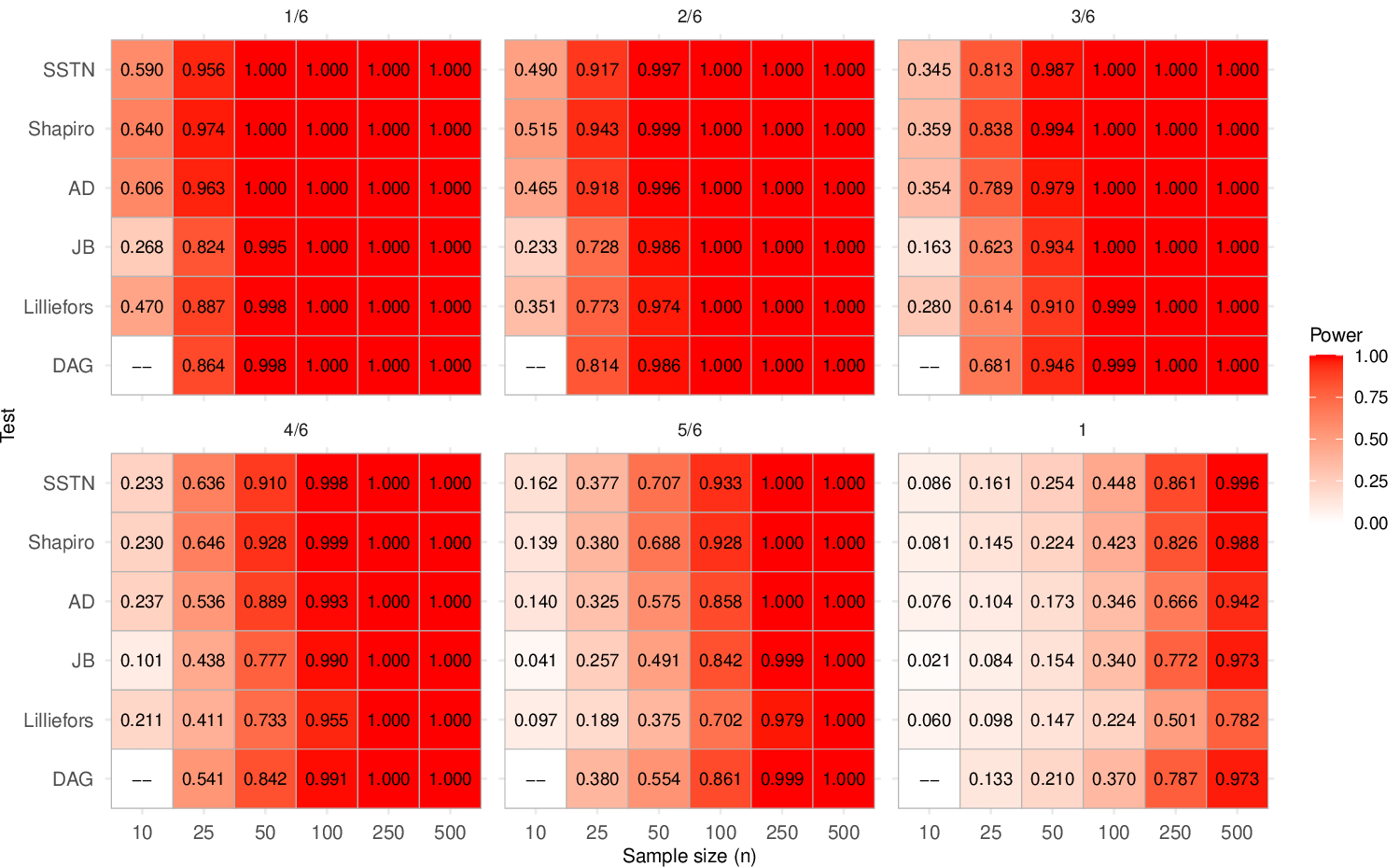}}
 \caption{Heatmaps of statistical power under the lognormal distribution, evaluated across sample sizes and tests and separated by the underlying parameter values. Cases with $n = 10$ for the D’Agostino–Pearson test are omitted.}
\label{fig:Lognormal}
\end{figure}

\begin{figure}[H]
 \centerline{\includegraphics[width=5.75in]{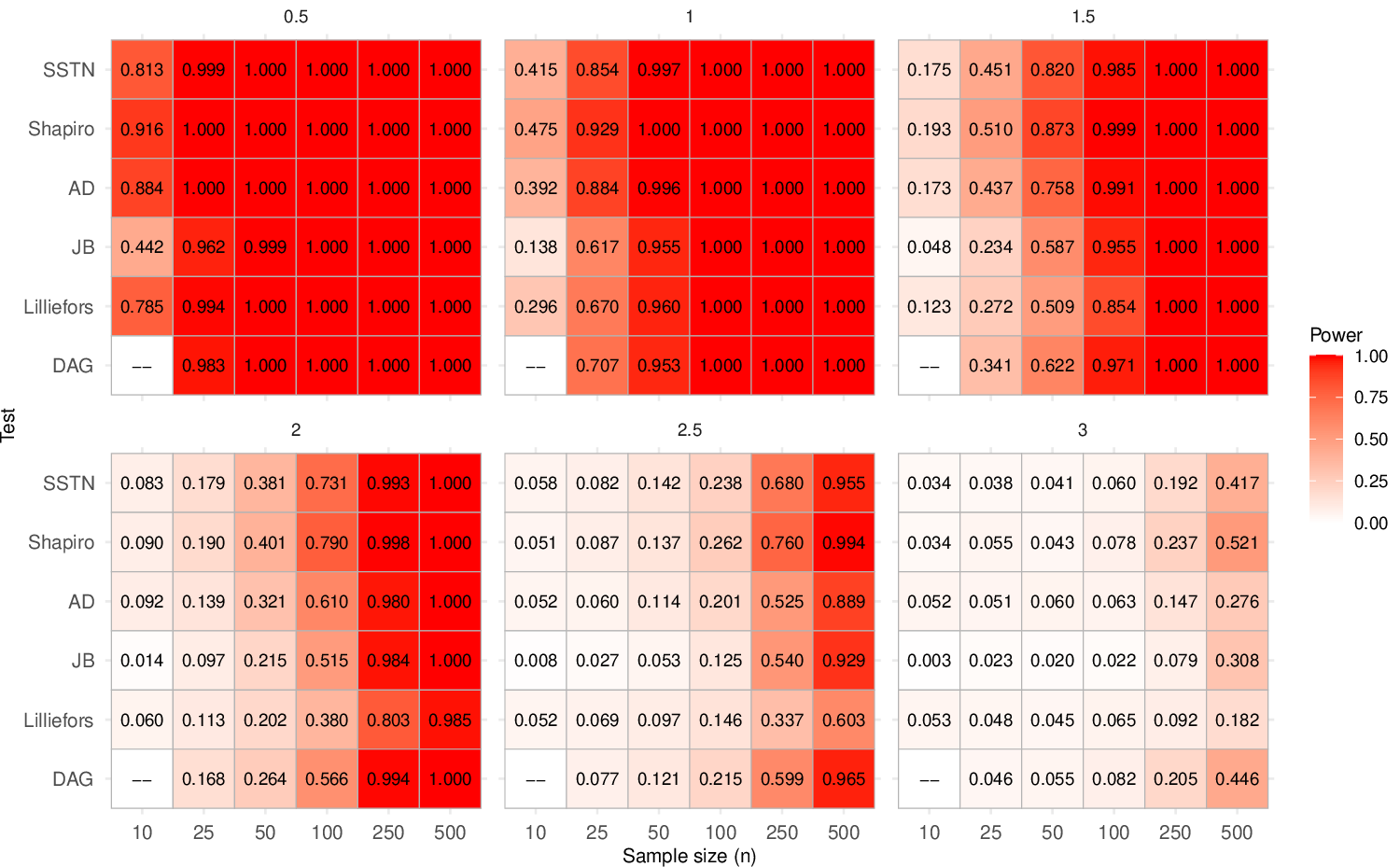}}
 \caption{Heatmaps of statistical power under the Weibull distribution, evaluated across sample sizes and tests and separated by the underlying parameter values. Cases with $n = 10$ for the D’Agostino–Pearson test are omitted.}
\label{fig:Weibull}
\end{figure}

\begin{figure}[H]
\vspace{1cm}
 \centerline{\includegraphics[width=5.75in]{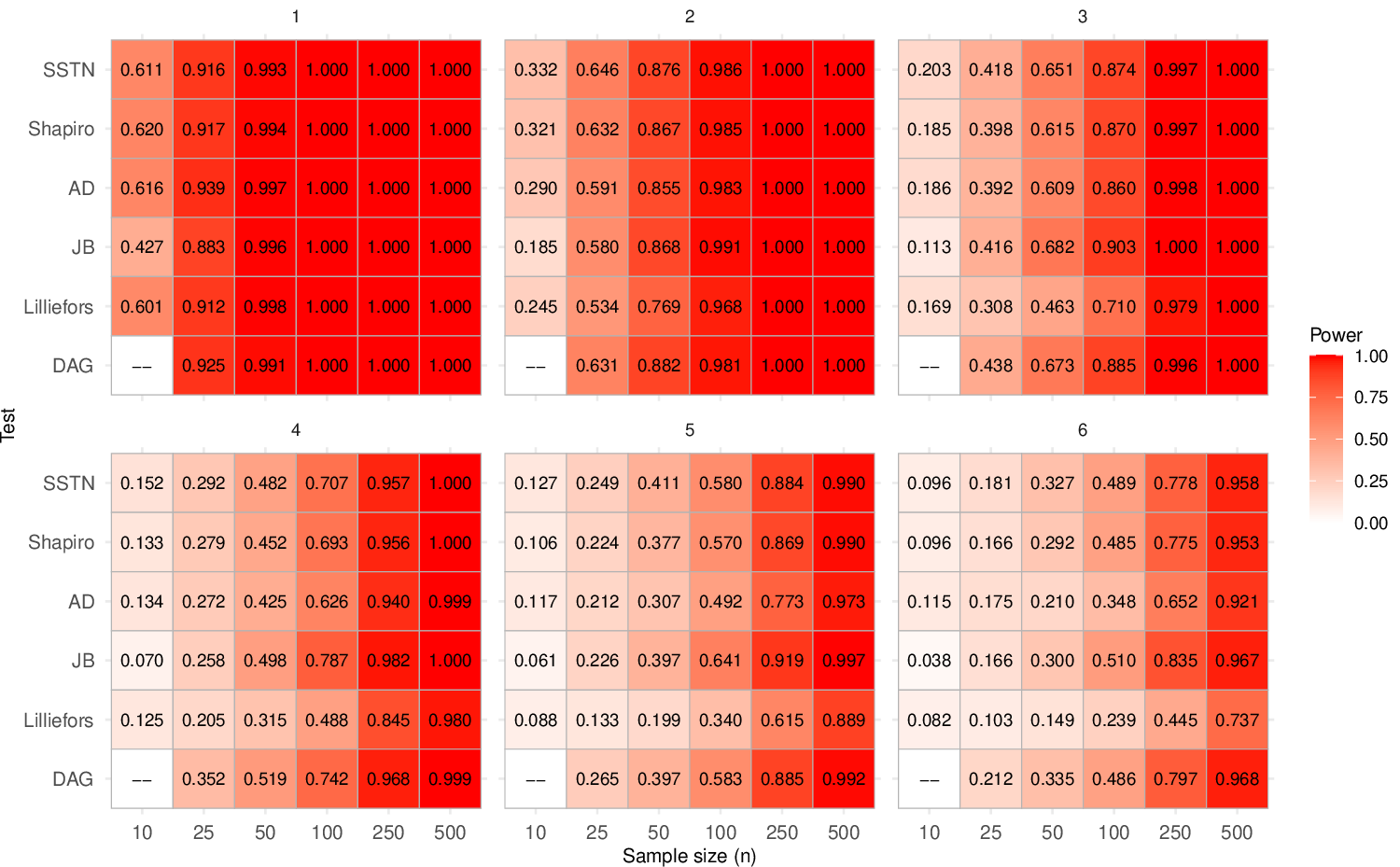}}
 \caption{Heatmaps of statistical power under the Student's $t$ distribution, evaluated across sample sizes and tests and separated by the underlying parameter values. Cases with $n = 10$ for the D’Agostino–Pearson test are omitted.}
\label{fig:t}
\end{figure}

\begin{figure}[H]
 \centerline{\includegraphics[width=5.75in]{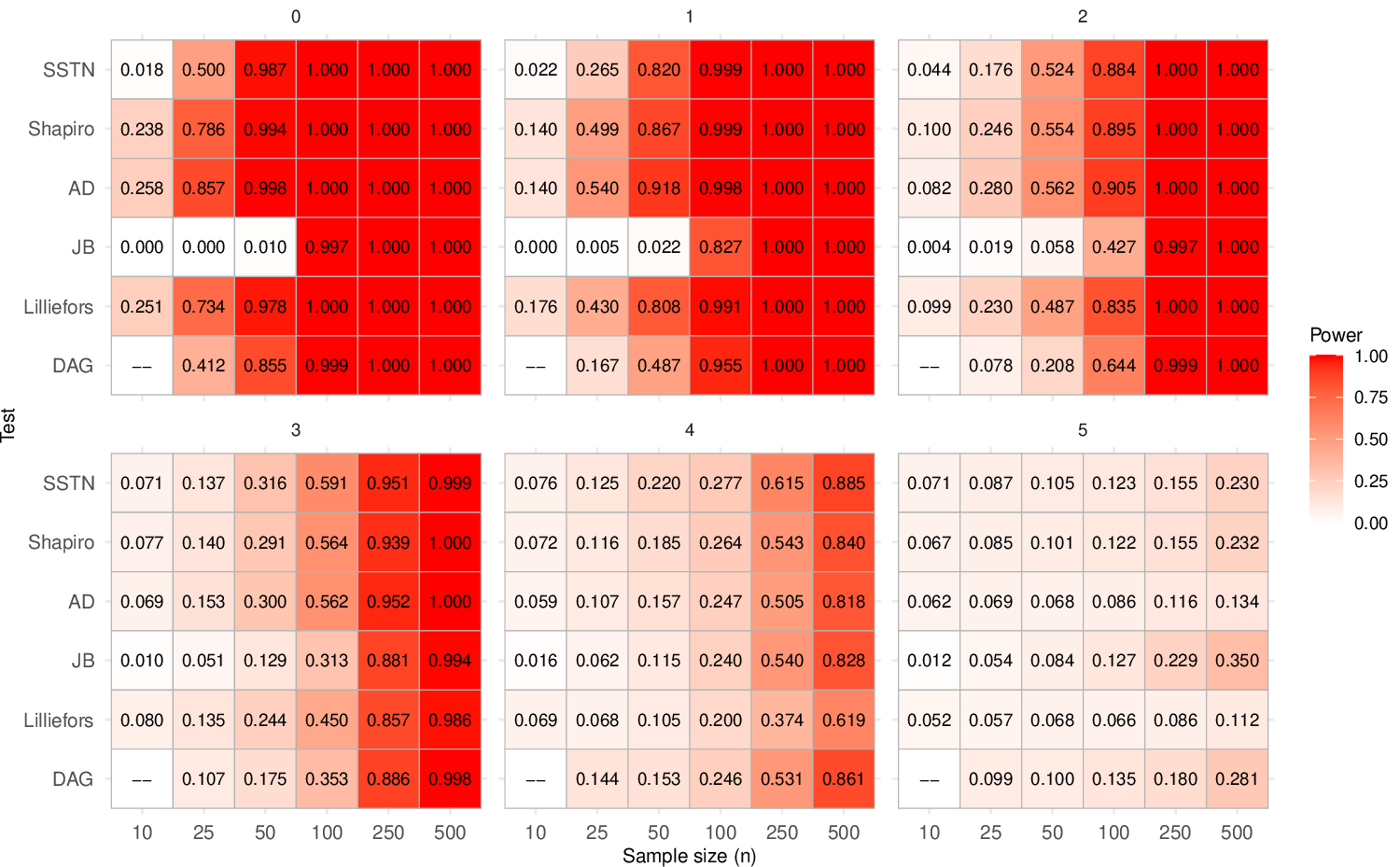}}
 \caption{Heatmaps of statistical power under the normal mixture model, evaluated across sample sizes and tests and separated by the underlying parameter values. Cases with $n = 10$ for the D’Agostino–Pearson test are omitted.}
\label{fig:Normal_mixture}
\end{figure}

\begin{figure}[H]
\vspace{1cm}
 \centerline{\includegraphics[width=5.75in]{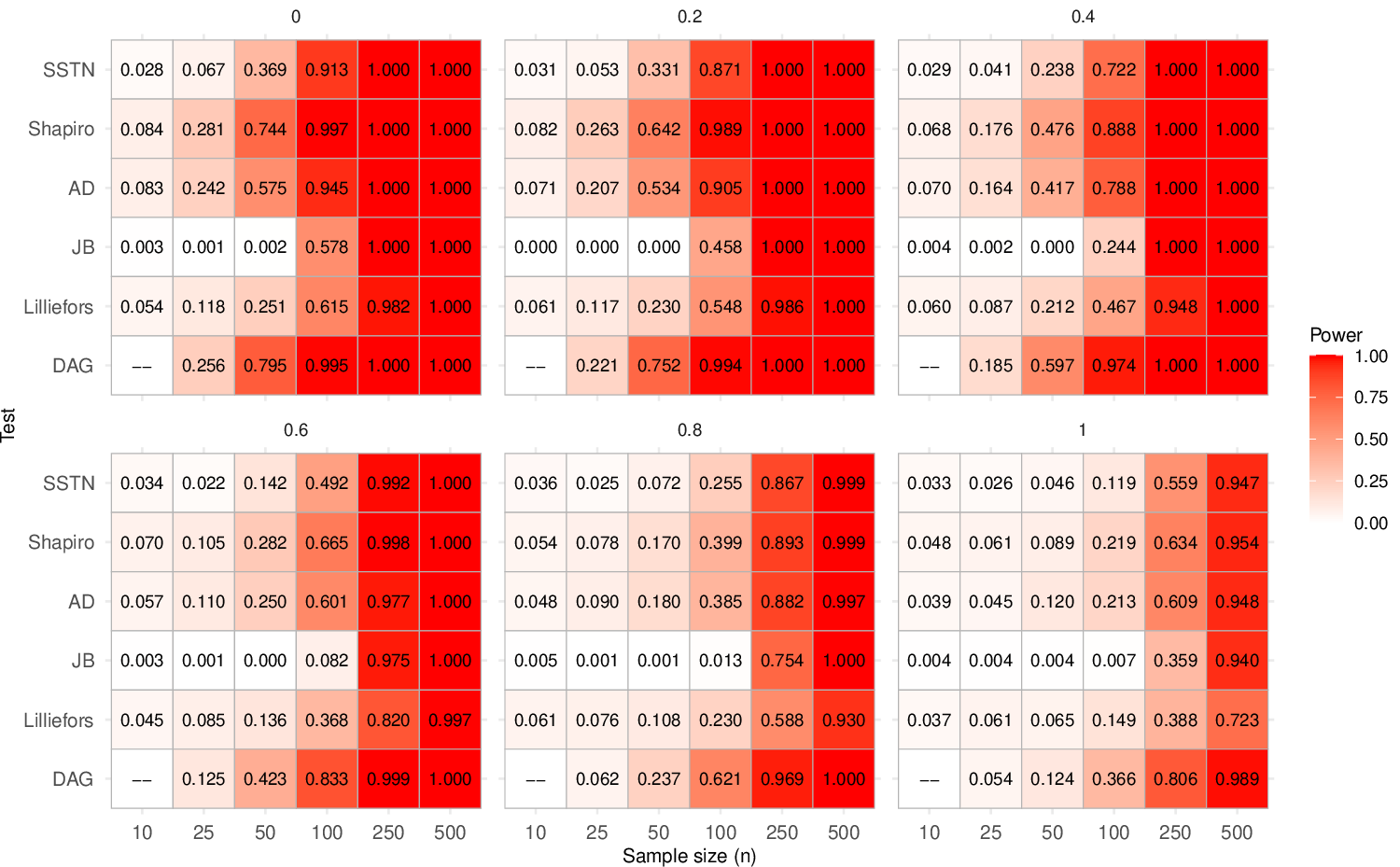}}
 \caption{Heatmaps of statistical power under the convolution of a uniform with a normal distribution, evaluated across sample sizes and tests and separated by the underlying parameter values. Cases with $n = 10$ for the D’Agostino–Pearson test are omitted.}
\label{fig:Uniform}
\end{figure}

\end{appendix}
\end{document}